\documentclass{msdm2014}
\usepackage{graphicx}
\usepackage{times}
\usepackage{latexsym}
\usepackage{helvet}
\usepackage{courier}
\usepackage{amsmath}
\usepackage{amssymb}
\usepackage{ntheorem}
\usepackage{url}
\usepackage[noend]{algorithmic}
\usepackage{algorithm}
\usepackage{multirow}
\usepackage{wrapfig}

\newtheorem*{prop-n}{Proposition}

\newtheorem{obs}{Observation}
\newtheorem*{obs-n}{Observation}

\newcommand{\oo}{\vec{o}}

\newenvironment{aroundtbl}{\begin{center}}{\end{center}}
  {\end{arroundtbl}
   \end{minipage}
   \end{center}}
\begin{document}
%
\title{Team Behavior in Interactive Dynamic Influence Diagrams with Applications to Ad Hoc Teams}

\numberofauthors{3}


\author{
%
\alignauthor
Muthukumaran Chandrasekaran\\
       \affaddr{THINC Lab}\\
       \affaddr{University of Georgia}\\
       \affaddr{Athens, GA, USA 30602}\\
       \email{mkran@uga.edu}
\alignauthor
Prashant Doshi\\
       \affaddr{THINC Lab}\\
       \affaddr{University of Georgia}\\
       \affaddr{Athens, GA, USA 30602}\\
       \email{pdoshi@cs.uga.edu}
\alignauthor
Yifeng Zeng\\
       \affaddr{School of Computing}\\
       \affaddr{Teesside University}\\
       \affaddr{Middlesbrough, Tees Valley, UK TS13BA}\\
       \email{y.zeng@tees.ac.uk}
\and
\alignauthor 
Yingke Chen\\
       \affaddr{Dept. of Computer Science}\\
       \affaddr{Queen's University Belfast}\\
       \affaddr{Belfast, Northern Ireland, UK}\\
       \email{y.chen@qub.ac.uk}
}
%





\maketitle

\begin{abstract}
Planning for ad hoc teamwork is challenging because it involves agents collaborating without any prior coordination or communication. The focus is on principled methods for a single agent to cooperate with others. This motivates investigating the ad hoc teamwork problem in the context of individual decision making frameworks. However, individual decision making in multiagent settings faces the task of having to reason about other agents' actions, which in turn involves reasoning about others. An established approximation that operationalizes this approach is to bound the infinite nesting from below by introducing level 0 models. We show that a consequence of the finitely-nested modeling is that we may not obtain optimal team solutions in cooperative settings. We address this limitation by including models at level 0 whose solutions involve learning. We demonstrate that the learning integrated into planning in the context of interactive dynamic influence diagrams facilitates optimal team behavior, and is applicable to ad hoc teamwork. 

\end{abstract}

, {Multiagent systems}



%

\section{Introduction}

Ad hoc teamwork involves a team of agents coming together to cooperate without any prior coordination or communication protocols~\cite{Stone10-adhoc}. The preclusion of prior commonality makes planning in ad hoc settings challenging. For example, well-known cooperative planning frameworks such as the Communicative multiagent team decision problem~\cite{Pynadath02:ComMTDP} and the decentralized partially observable Markov decision process (DEC-POMDP)~\cite{Bernstein02:Complexity} utilize centralized planning and distribution of local policies among agents, which are assumed to have common initial beliefs. These assumptions make the frameworks unsuitable for ad hoc teamwork.     

A focus on how an agent should behave online as an ad hoc teammate informs previous approaches toward planning. This includes an algorithm for online planning in ad hoc teams (OPAT)~\cite{WZCijcai11} that solves a series of stage games assuming that other agents are optimal with the utility at each stage computed using Monte Carlo tree search. Albrecht and Ramamoorthy~\cite{Stefano13HBA} model the uncertainty about other agents' types and construct a Harsanyi-Bayesian ad hoc game that is solved online using learning. While these approaches take important steps, they assume that the physical states and actions of others are perfectly observable, which often may not apply.

The focus on individual agents' behaviors in ad hoc teams motivates that we situate the problem in the context of individual decision-making frameworks.
In this regard, recognized frameworks include the interactive POMDP (I-POMDP)~\cite{Gmytrasiewicz05:Framework:JAIR}, its graphical counterpart, interactive dynamic influence diagram (I-DID)~\cite{Doshi09:Graphical}, and  I-POMDP Lite~\cite{Hoang13:iPOMDPLite}. These frameworks allow considerations of partial observability of the state and uncertainty over the types of other agents with minimal prior assumptions, at the expense of increased computational complexity. Indeed, Albrecht and Ramamoorthy~\cite{Stefano13HBA} note the suitability of these frameworks to the problem of ad hoc teamwork but find the complexity challenging.  

While recent advances on model equivalence~\cite{Zeng12:Exploiting} allow frameworks such as I-DIDs to scale, another significant challenge that merits attention is due to the finitely-nested modeling used in these frameworks, which assumes the presence of level 0 models that do not explicitly reason about others~\cite{Brandenburger93:Hierarchies,Aumann99:Interactive2,camerer2004cognitive,Mertens85:Formulation}. A consequence of this approximation is that we may not obtain optimal solutions in cooperative settings. To address this, we augment the I-DID framework by additionally attributing a new type of level 0  model.  This  type distinguishes  itself  by  utilizing reinforcement learning (RL) either online or in simulation to discover possible collaborative policies that a level 0 agent may execute.

The contributions of this paper are two-fold: First, we show the emergence of true team behavior when the reasoning ability of lower level agents is enhanced via learning. We demonstrate globally optimal teammate solutions when agents are modeled in finitely-nested \textit{augmented} I-DIDs (Aug. I-DIDs) while \textit{traditional} I-DIDs fail. Second, we demonstrate the applicability of Aug. I-DIDs to ad hoc settings and show its effectiveness for varying types of teammates. For this, we utilize the ad hoc setting of Wu et al.~\cite{WZCijcai11}, and experiment  with  multiple well-known  cooperative domains. We also perform a baseline comparison of Aug. I-DIDs with an implementation of a generalized version of OPAT that accounts for the partial observability.

\section{Background: Interactive DIDs}
\label{sec:background}

We      sketch      I-DIDs      below      and      refer      readers
to~\cite{Zeng12:Exploiting} for more details.

\subsection{Representation}
\begin{figure}[!ht]
\begin{center}
\begin{minipage}{2.7in}
\centerline{\includegraphics[width=2.7in]{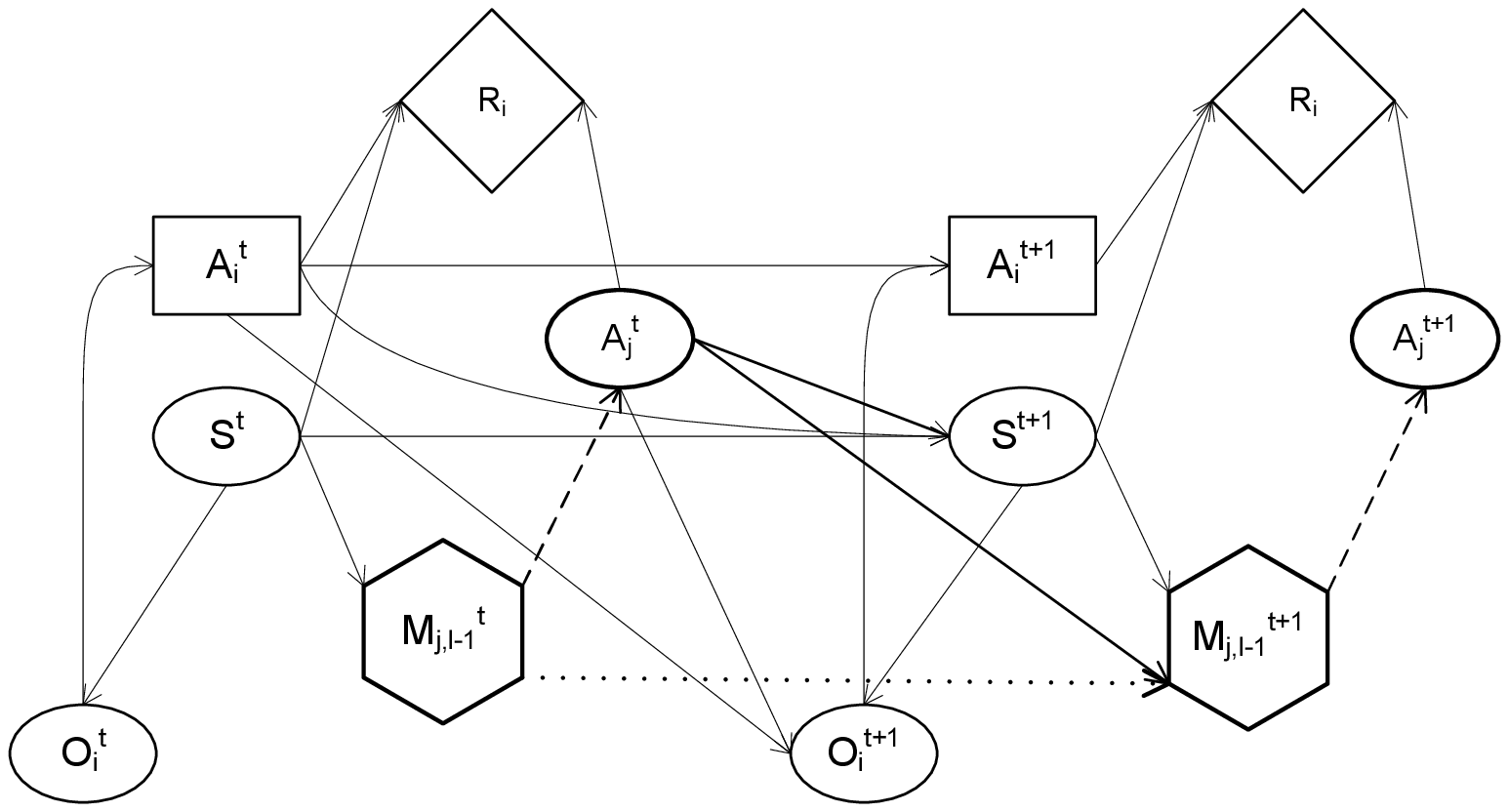}}
\centerline{{\small $(a)$}}
\end{minipage}
\begin{minipage}{2.7in}
\centerline{\includegraphics[width=2.7in]{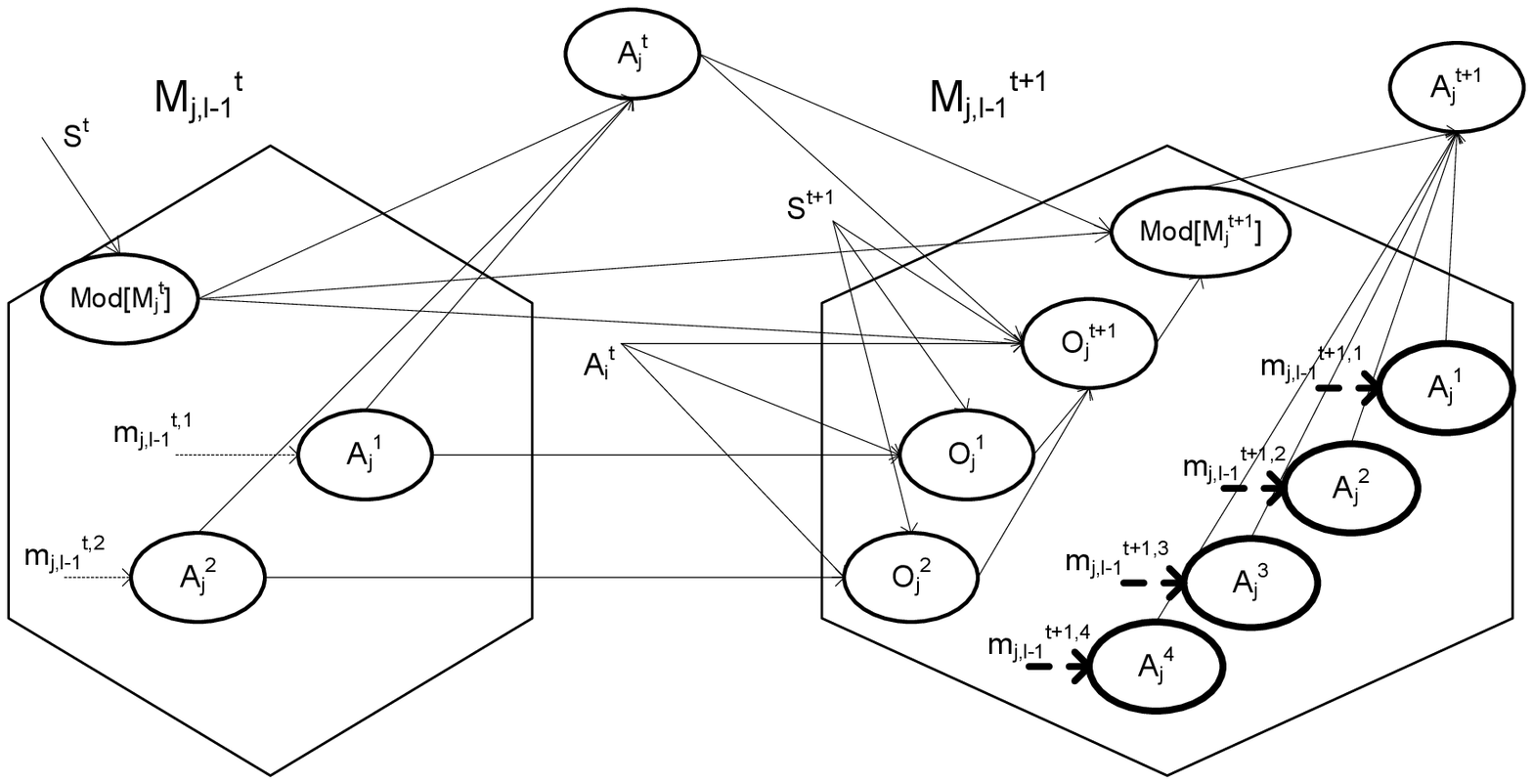}}
\centerline{{\small $(b)$}}
\end{minipage}
\end{center}
\vspace{-0.1in}
\caption{{\small $(a)$ A generic two time-slice level $l$ I-DID for agent $i$.
  The dotted model  update link represents the update  of $j$'s models
  and the distribution over the models over time; $(b)$ Implementation
  of the model update link  using standard dependency links and chance
  nodes;  e.g., two  models, $m_{j,l-1}^{t,1}$  and $m_{j,l-1}^{t,2}$,
  are updated into four models~(shown in bold) at time $t+1$.}}
\label{fig:I-DID}
\vspace{-0.1in}
\end{figure}

A traditional DID models sequential  decision making for a single agent by
linking a set of chance, decision and utility nodes over multiple time
steps.  To  consider multiagent  interactions, I-DIDs introduce  a new
type of node called the {\em model node}~(hexagonal node, $M_{j,l-1}$,
in  Fig.~\ref{fig:I-DID}($a$)) that represents  how another  agent $j$
acts as the subject agent  $i$ reasons about  its own decisions  at level
$l$.  The model node contains a set of $j$'s candidate models at level
$l-1$ ascribed by $i$.  A link from the chance node, $S$, to the model
node, $M_{j,l-1}$,  represents agent $i$'s beliefs  over $j$'s models.
Specifically,  it is  a  probability distribution  in the  conditional
probability   table   (CPT)  of   the   chance  node,   $Mod[M_j]$~(in
Fig.~\ref{fig:I-DID}($b$)).  An individual  model of $j$, $m_{j,l-1} =
\langle b_{j,l-1}, \hat{\theta}_{j} \rangle$, where $b_{j,l-1}$ is the
level $l-1$ belief, and  $\hat{\theta}_{j}$ is the agent's {\em frame}
encompassing the decision, observation  and utility nodes. Each model,
$m_{j,l-1}$, could  be either a  level $l-1$ I-DID  or a DID  at level
0. Solutions  to the model are  the predicted behavior of  $j$ and are
encoded into the  chance node, $A_j$, through a  dashed link, called a
{\em policy  link}.  Connecting  $A_j$ with other  nodes in  the I-DID
structures  how agent  $j$'s actions  influence  $i$'s decision-making
process.

Expansion of an I-DID involves the  update of the model node over time
as  indicated by  the {\em  model update  link} -  a dotted  link from
$M_{j,l-1}^t$  to $M_{j,l-1}^{t+1}$  in  Fig.~\ref{fig:I-DID}($a$). As
agent $j$ acts and receives  observations over time, its models should
be  updated.   For  each  model
$m_{j,l-1}^t$  at time  $t$,  its optimal  solutions  may include  all
actions and  agent $j$ may  receive any of the  possible observations.
Consequently, the set of the  updated models at $t+1$ contains up
to     $|\mathcal{M}_{j,l-1}^t||A_j||\Omega_j|$     models.      Here,
$|\mathcal{M}_{j,l-1}^t|$ is  the number of  models at time $t$,
and  $|A_j|$  and  $|\Omega_j|$  the  largest spaces  of  actions  and
observations respectively among all the models. 
The   CPT   of    $Mod[M_{j,l-1}^{t+1}]$   specifies   the   function,
$\tau(b_{j,l-1}^t,a_j^t,o_j^{t+1},b_{j,l-1}^{t+1})$ which  is 1 if the
belief  $b_{j,l-1}^t$  in the  model  $m_{j,l-1}^t$  using the  action
$a_j^t$ and observation $o_j^{t+1}$  updates to $b_{j,l-1}^{t+1}$ in a
model  $m_{j,l-1}^{t+1}$; otherwise,  it is  0. We  may  implement the
model update link using standard dependency links and chance nodes, as
shown  in Fig.~\ref{fig:I-DID}($b$),  and  transform an  I-DID into  a
traditional  DID.

\subsection{Solution}

A level $l$ I-DID of agent  $i$ expanded over $T$ time steps is solved
in a bottom-up manner. To solve agent $i$'s level $l$ I-DID, all lower
level $l-1$ models  of agent $j$ must be solved.   Solution to a level
$l-1$ model, $m_{j,l-1}$, is $j$'s policy that is a mapping from $j$'s
observations  in $O_j$  to the  optimal  decision in  $A_j$ given  its
belief, $b_{j,l-1}$.   Subsequently, we may  enter $j$'s
optimal decisions into  the chance node, $A_j$, at  each time step and
expand $j$'s models in $Mod[M_{j,l-1}]$  corresponding to each pair of $j$'s
optimal action and  observation.  We perform this process  for each of
level $l-1$  models of  $j$ at  each time step,  and obtain  the fully
expanded level $l$ model.
We outline the
algorithm  for exactly  solving I-DIDs  in  Fig.~\ref{fig:ExAlgo}.

The computational complexity of solving I-DIDs is mainly due to the exponential growth of lower $l$-1 $j$'s models over time. Although the  space of possible models  is very large,  not all models
need  to  be considered  in  the model  node.   Models  that are  behaviorally equivalent (BE)~\cite{Pynadath07:Minimal}
-- whose behavioral  predictions for the other agent  are identical --
could be pruned and a single representative model can be considered.  This is
because the solution  of the subject agent's I-DID  is affected by the
  behavior  of  the  other  agent  only;  thus  we  need  not
distinguish   between  BE   models.  Let   {\bf
  PruneBehavioralEq} ($\mathcal{M}_{j,l-1}$)  be the  procedure  that prunes
BE   models  from   $\mathcal{M}_{j,l-1}$
returning the representative models~(line 6).

 Note  that  lines  4-5~(in  Fig.~\ref{fig:ExAlgo}) solve  level  $l$-1
 I-DIDs or DIDs and then supply the policies to level $l$ I-DID. Due to
 the bounded rationality of level $l$-1 agents, the solutions lead to a
 suboptimal  policy of  agent  $j$, which  certainly compromises  agent
 $i$'s  performance   in  the  interactions  particularly   in  a  team
 setting. Also, note that the level 0 models are DIDs that do not involve learning. We will show in the coming sections that solving I-DIDs integrated with
 RL may generate the expected team behavior among agents $i$ and $j$.
\noindent
\newcounter{ln}
\setcounter{ln}{0}
\begin{figure}[ht!]
\framebox[3.3in]{
\begin{minipage}{3.2in}
\small{
\begin{tabbing}
{\sc \textbf{I-DID Exact}}(level $l \geq 1$ I-DID or level 0 DID, horizon $T$)\\
\underline{Expansion Phase}\\
\stepcounter{ln} \theln. {\bf For} \=  $t$ {\bf from} 0 {\bf to} $T-1$ {\bf do} \+\\
\<\stepcounter{ln} \theln.\> {\bf If} \= $l \geq 1$ {\bf then}\+\\
{\em \underline{Populate $M_{j,l-1}^{t+1}$}}\\
\<\<\stepcounter{ln} \theln.\>\> {\bf For} \= {\bf each} $m_j^t$ {\bf in} $\mathcal{M}_{j,l-1}^t$ {\bf do} \+\\
\<\<\<\stepcounter{ln} \theln.\>\>\> Recursively call algorithm with the $l-1$ I-DID\\ (or DID) that represents $m_j^t$ and horizon, $T-t$\\
\<\<\<\stepcounter{ln} \theln.\>\>\> Map the decision node of the solved I-DID (or DID),\\ $OPT(m_j^t)$, to the corresponding chance node $A_j$\-\\
\<\<\stepcounter{ln} \theln.\>\> $\mathcal{M}_{j,l-1}^{t}$ $\leftarrow$ {\bf PruneBehavioralEq}($\mathcal{M}_{j,l-1}^{t}$)\\
\<\<\stepcounter{ln} \theln.\>\> {\bf For} \= {\bf each} $m_j^t$ {\bf in} $\mathcal{M}_{j,l-1}^t$ {\bf do} \+\\
\<\<\<\stepcounter{ln} \theln.\>\>\> {\bf For} \= {\bf each} $a_j$ {\bf in} $OPT(m_j^t)$ {\bf do} \+\\
\<\<\<\<\stepcounter{ln} \theln.\>\>\>\> {\bf For} \= {\bf each} $o_j$ {\bf in} $O_j$ (part of $m_j^t$) {\bf do} \+\\
\<\<\<\<\<\stepcounter{ln} \theln.\>\>\>\>\> Update $j$'s belief, $b_j^{t+1} \leftarrow SE(b_j^t,a_j,o_j)$\\
\<\<\<\<\<\stepcounter{ln} \theln.\>\>\>\>\> $m_j^{t+1}$ $\leftarrow$  New I-DID (or DID) with $b_j^{t+1}$\\
\<\<\<\<\<\stepcounter{ln} \theln.\>\>\>\>\> $\mathcal{M}_{j,l-1}^{t+1} \overset{\cup}{\leftarrow} \{m_j^{t+1}\}$\-\-\-\\
\<\<\stepcounter{ln} \theln.\>\> Add the model node, $M_{j,l-1}^{t+1}$, and the model update link\-\\
\<\stepcounter{ln} \theln.\> Add the chance, decision, and utility nodes for\\ $t+1$ time slice and the dependency links between them\\
\<\stepcounter{ln} \theln.\> Establish the CPTs for each chance node and utility node\-\\
\underline{Solution Phase}\\

\stepcounter{ln} \theln. {\bf If} \= $l \geq 1$ {\bf then}\+\\
\<\stepcounter{ln} \theln.\> Represent the model nodes, policy links and the model\\ update links as in Fig.~\ref{fig:I-DID} to obtain the DID \-\\
\stepcounter{ln} \theln. Apply the standard look-ahead and backup
method\\~~~~~~ to solve the expanded DID
\end{tabbing}}
\end{minipage}}
\caption{{ \small Algorithm for  exactly solving  a level $l  \geq 1$  I-DID or
  level 0 DID  expanded over $T$ time steps.}}
\label{fig:ExAlgo}
\vspace{-0.1in}
\end{figure}

\section{Teamwork in Interactive DIDs}
\label{sec:teamwork}

Ad hoc teamwork involves multiple  agents working collaboratively in
order to  optimize the  team reward.  Each ad hoc agent in the  team behaves
according to a  policy, which maps the agent's  observation history or
beliefs to the  action(s) it should perform. We  begin by showing that
the  finitely-nested  hierarchy in  I-DID  does  not facilitate  ad hoc teamwork. However, augmenting the  traditional model space with models
whose solution is obtained via reinforcement learning  provides a way for team behavior to
emerge.

\subsection{Implausibility of Teamwork}

\begin{figure}[!ht]
  \centering
    \includegraphics[width=1.7in]{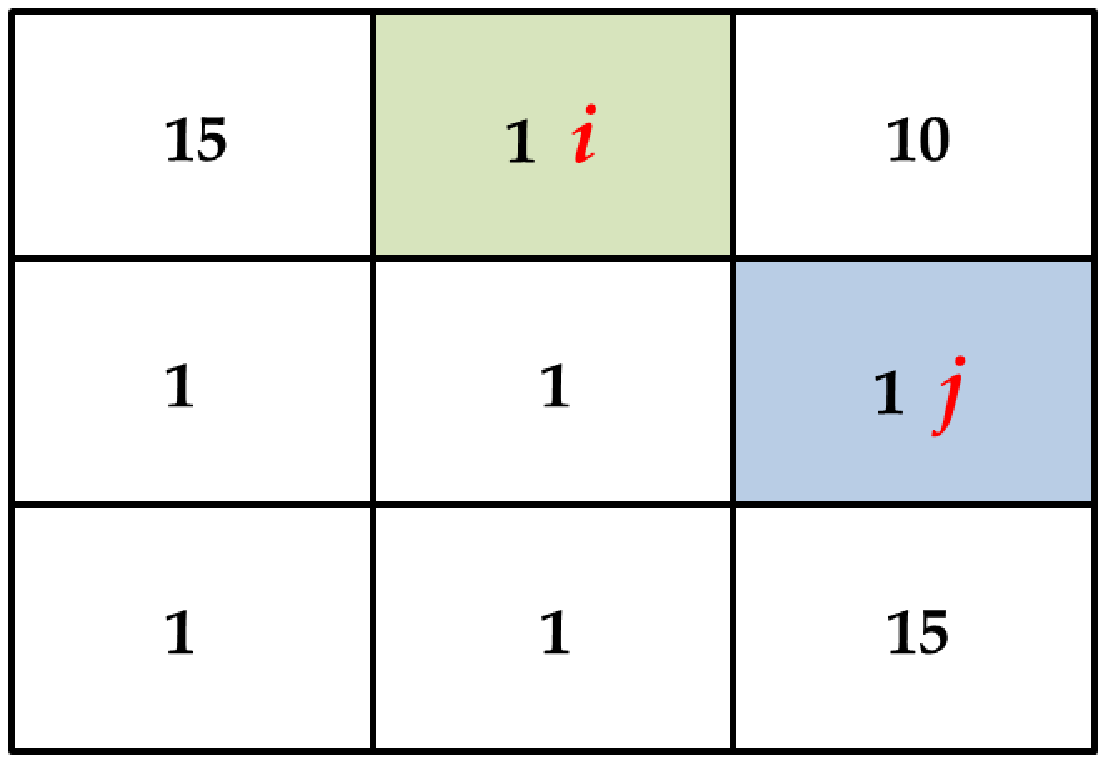}
  \caption{{\small Agents $i$  and $j$  in the grid  meeting problem  with the
    numbers being  their individual rewards.}}
\label{fig:grid}
\end{figure}

Fig.~\ref{fig:grid} shows an ad hoc team setting of a two-agent {\em grid meeting}
problem~\cite{Bernstein05:Bounded}. The  agents can detect the presence of a wall  on its right ($RW$), left ($LW$) or the absence of it on both sides ($NW$).  Given a specific observation, the agent may  choose to  either move  in one of  four directions  -- south  ($MS$), north  ($MN$),  east  ($ME$),  or west ($MW$), or  stay in the same  cell ($ST$). Each ad hoc agent, $i$ or $j$, moves in the grid and collects rewards as the number indicated in the occupied cell. If they move to different cells, the agents get their own individual reward. However, if they move to the same cell allowing them to hold an ad hoc meeting, they will be rewarded with twice the sum of their individual rewards. Initial positions of  the two agents are  shown in color and we  focus on their immediate actions.

If  each agent deliberates at its  own level, agent
$i$ modeled at level 0 will choose  to move left while a level 0 agent
$j$ chooses to move down. Each  agent will obtain a reward of 15 while
the whole team  gets 30. Agent $i$ modeled at  level 1 and modeling
$j$  at level  0 thinks  that $j$  will move  down, and  its  own best
response to predicted  $j$'s behavior is to move  left. Analogously, a
level 1 agent $j$ would choose to move down.  A level 2 agent $i$ will
predict that a level 1 $j$  moves down as mentioned previously, due to
which  it decides  to move  left.  Analogously,  a level  2  agent $j$
continues  to  decide to  move  down.   We  may apply  this  reasoning
inductively to conclude  that level $l$ agents $i$  and $j$ would move
left and  down, respectively,  thereby earning a  joint reward  of 30.
However, the optimal team behavior in  this setting is for $i$ to move
right  and  $j$  to  move  up  thereby  obtaining  a  team  reward  of
40. 

Clearly, these finite hierarchical systems preclude the agents' optimal teamwork due to the bounded reasoning of the lowest level (level  0) agents.  The following \textit{observation} states this more formally:

\begin{obs}
 There  exist cooperative  multiagent settings  in  which rational intentional
  agents each modeled using the finitely-nested I-DID (or I-POMDP) may not choose  the jointly optimal behavior  of working
  together as a team.
\label{obs:paradox}
\end{obs}

Notice that an offline specification of level 0 models in cooperative
settings is necessarily incomplete.   This is because the true benefit
of cooperative  actions often  hinges on others  performing supporting
actions, which by themselves may not  be highly rewarding
to the agent. Thus, despite solving the level 0 models optimally,
the agent may not engage in optimal team behavior.

In general, this observation holds for cooperative settings where the self-maximizing level 0  models result in predictions that are not consistent with team behavior. Of course, settings may exist where the level 0 model's solution coincides with the policy  of a teammate thereby leading to joint teamwork. Nevertheless, the significance of this observation is that we cannot  rely on finitely-nested I-DIDs to generate optimal teammate policies.


We observe that  team behavior is challenging in  the context we study
above because of the bounded rationality imposed by assuming the existence of a level 0. The boundedness precludes  modeling others at the same  level as one's
own  --  as  an equal  teammate.   However,  at  the same  time,  this
imposition  is, $(a)$  motivated  by reasons  of computability,  which
allow us  to operationalize  such a paradigm;  and $(b)$ allows  us to
avoid some self-contradicting, and therefore impossible beliefs, which
exist       when      infinite       belief       hierarchies      are
considered~\cite{Binmore90:Essays}.
Consequently, this work is of significance because it  may provide us
a way  of generating  optimal team  behavior  in finitely-nested
frameworks, which so far have been utilized for noncooperative
settings, and provides a principled way to solving ad hoc teamwork problems.

\subsection{Augmented Level 0 Models that Learn}

We present a principled way to induce team behavior by enhancing the reasoning ability of lower-level agents.
While  it is difficult to {\em  a priori} discern the benefit of moving up for agent $j$ in Fig.~\ref{fig:grid}, it could be {\em experienced}  by the agent.  Specifically, it  may explore moving in  different  directions including  moving  up  and  learn about  its benefit from  the ensuing, possibly indirect, team  reward.

Subsequently, we may expect an agent to learn policies that  are consistent  with optimal teammate  behavior because the corresponding  actions provide large  reinforcements. For example, given that agent $i$ moves right in Fig.~\ref{fig:grid}, $j$ may choose to move up in its exploration, and thereby  receive a  large reinforcing  reward.  This observation  motivates formulating  level 0 models that utilize  RL to generate  the predicted policy for the modeled agent. Essentially, we expect that RL with its explorations would compensate for the lack of teamwork caused by bounded reasoning in finitely-nested I-DIDs.  

Because the  level 0  models generate policies  for the  modeled agent only, we focus  on the modeled agent's learning problem. However,  the rewards in the multiagent setting usually depend  on actions of all agents  due to which the other agent  must be simulated  as well.  The other  agent's actions are a part of  the environment and its presence  hidden at level 0,  thereby  making  the problem  one  of  single-agent learning as opposed to one of multi-agent learning.



We  augment  the level  0  model  space,  denoted as $\mathcal{M}'_{j,0}$, by  additionally attributing  a new  type of  level  0 model  to the  other agent  $j$:
$m'_{j,0} = \langle  b_{j,0}, \hat{\theta}'_j\rangle$, where $b_{j,0}$
is $j$'s belief and $\hat{\theta}'_{j,0}$ is the frame of the learning
model. The frame, $\hat{\theta}'_{j,0}$, consists of the learning rate,
$\alpha$;  a seed policy,  $\pi'_j$, of  planning horizon,  $T$, which
includes  a fair  amount of  exploration; and  the chance  and utility
nodes of the DID  along with a candidate policy of agent $i$, which could be an arbitrary policy from $i$'s policy space, $\Pi_i$, as agent $i$'s actual behavior is not known. This permits a proper simulation of the environment.


This type of model, $m'_{j,0}$, differs from a traditional DID based level
0 model  in the aspect that  $m'_{j,0}$ does not  describe the offline
planning process of how agent  $j$ optimizes its decisions, but allows
$j$  to learn  an optimal  policy,  $\pi_j$, with  the learning  rate,
either online or in a simulated setting. Different models of agent $j$ differ not only in their learning rates and seed policies, but also in the $i$'s candidate policy that is used. In principle, while the learning rate and seed policies may be held fixed, $j$'s model space could be as large as $i$'s policy space. Consequently, our augmented model space becomes extremely large.

\subsection{Model-Free Learning: Generalized MCESP}

Learning has  been applied to  solve decision-making problems  in both
single-      and       multi-agent      settings.       Both      model
based~\cite{Chrisman92:RLP} and model free~\cite{Meuleau99learningfinite-state,Perk02:RLPOMDP}   learning
approaches   exist    for   solving   POMDPs. In the multiagent context,    Banerjee   {\em   et al.}~\cite{Banerjee12:DRL}   utilized  a  semi-model based distributed   RL  for  finite   horizon   DEC-POMDPs.    Recently,   Ng   {\em   et  al.}~\cite{Ng12:BAIPOMDP} incorporated model learning in the context of I-POMDPs where the subject agent learns  the transition and  observation probabilities  by augmenting  the  interactive states with frequency counts.

Because the  setting in  which the learning  takes place  is partially
observable,  RL  approaches  that   compute  a  table  of  values  for
state-action  pairs  do not  apply. We  adapt Perkin's Monte  Carlo Exploring Starts  for POMDPs (MCESP)~\cite{Perk02:RLPOMDP},  which has
been shown to learn good policies in fewer iterations while making no prior assumptions about the agent's models in order to achieve convergence.  MCESP maintains
a $Q$  table indexed  by observation, $o_j$,  and action,  $a_j$, that
gives  the   value  of  following  a  policy,   $\pi_j$,  except  when
observation,  $o_j$,  is obtained  at  which  point   action,  $a_j$,  is
performed.   An agent's policy  in MCESP  maps a single  observation to
actions  over  $T$  time   horizons.   We  generalize  MCESP  so  that
observation  histories of  length up  to  $T$, denoted  as $\oo$,  are
mapped  to actions.   A table  entry, $Q_{\oo,a}^{\pi_j}$,  is updated
over every  simulated trajectory  of agent $j$,  $\tau$ $=$  $\{$ $*$,
$a_j^0$,  $r_j^0$, $o_j^1$,  $a_j^1$, $r_j^1$,  $\cdots$, $o_j^{T-1}$,
$a_j^{T-1}$, $r_j^{T-1}$, $o_j^T$ $\}$, where $r_j$ is the team reward
received. Specifically, the $Q_{\oo,a}^{\pi_j}$ value is updated as:
\begin{equation}
Q_{\oo,a}^{\pi_j}\leftarrow(1-\alpha)Q_{\oo,a}^{\pi_j} + \alpha R_{post-\oo}(\tau)
\label{Eq:QValue}
\end{equation}
where $\alpha$  is the learning  rate and $R_{post-\oo}(\tau)$  is the
sum  of  rewards of  a  portion  of  the observation-action  sequence,
$\tau$,  following the  first occurrence  of $\oo$  in $\tau$,  say at
$t'$:  $R_{post-\oo}(\tau)$=$\sum_{t=t'}^{T-1}  \gamma^t  r_t$,  where
$\gamma  \in [0,1)$ is  the discount  factor.  Alternate  policies are
considered by perturbing the  action for randomly selected observation
histories.

Level 0 agent $j$ learns its policy while agent $i$'s actions are hidden in the environment. In other words, agent $j$ needs to reason with unknown behavior of $i$ while it learns level 0 policy using the generalized MCESP algorithm. Agent $j$ considers the entire policy space of agent $i$, $\Pi_i$,  and a fixed policy of $i$, $\pi_i$($\in \Pi_i$), results in one learned $j$'s policy, $\pi_j$.

We show the algorithm for solving level 0 models using the generalized
MCESP  in Fig.~\ref{fig:RLALG}. The algorithm  takes as  input agent
$j$'s  model whose solution  is needed  and the  policy of  $i$, which
becomes a part of the  environment. We repeatedly obtain a trajectory,
$\tau$, of length $T$ either by running the agent online or simulating
the environment by sampling  the states, actions and observations from
the  appropriate  CPTs  (lines  5-10).   The  trajectory  is  used  in
evaluating  the value  of the  current policy,  $\pi_j$, of  agent $j$
(line 11).  Initially,  we utilize the seed policy  contained in agent
$j$'s model.   If another action, $a'$, for  the observation sequence,
$\oo$, is optimal, we update $\pi_j$ to conditionally use this action,
otherwise the policy remains unchanged (lines 12-13). This is followed
by generating a  perturbed policy in the neighborhood  of the previous
one  (line  14),  and  the  evaluation  cycle  is  repeated.   If  the
perturbation  is  discarded  several   times,  we  may  terminate  the
iterations and return the current policy.

\setcounter{ln}{0}
\begin{figure}[!ht]
\begin{small}
\framebox[3.3in]{
\begin{minipage}{3.3in}
\small{
\begin{tabbing}
{\sc RL for Level 0 Model} ($j$'s model,  $m'_{j,0}$, $i$'s policy,
$\pi_i$, $T$)\\\\

\stepcounter{ln} \theln. Sample the initial state $s$
from $b_{j,0}$ in $j$'s model\\
\stepcounter{ln} \theln. Set current policy of $j$ denoted as $\pi_j$ using the seed\\
~~~~~~policy in $j$'s model\\
\stepcounter{ln} \theln. Set $\tau \leftarrow \{*\}$  (empty observation)\\
\stepcounter{ln} \theln. {\bf Rep}\={\bf eat} \+\\
\< \stepcounter{ln} \theln.\> {\bf For} \= $t$ = 0 to $T-1$ {\bf do}\+\\
\<\<\stepcounter{ln} \theln.\>\> Obtain $i$'s action from $\pi_i$ and
$j$'s action, $a_j^t$, \\ from current policy of $j$ using observation
history\\
\<\<\stepcounter{ln} \theln.\>\> Obtain the next state, $s'$, either
by performing\\ the actions or sampling \\
\<\<\stepcounter{ln} \theln.\>\> Obtain team reward, $r_j^t$, using state and joint actions \\
\<\<\stepcounter{ln} \theln.\>\> Obtain $j$'s observation,
$o_j^{t+1}$, using next\\ state and joint actions\\
\<\<\stepcounter{ln} \theln.\>\> Generate trajectory, $\tau \leftarrow \tau \cup \{a_j^t,r_j^t,o_j^{t+1}\}$\-\\
\< \stepcounter{ln} \theln.\> Update $Q_{\oo,a}^{\pi_j}$ according to Eq.~\ref{Eq:QValue}\\
\< \stepcounter{ln} \theln.\>  {\bf If} \= $\underset{a'}{max}$ $Q_{\oo,a'}^{\pi_j}$ $>$ $Q_{\oo,a}^{\pi_j}$ \+\\
\<\<\stepcounter{ln} \theln.\>\> $\pi_j(\oo) \leftarrow a'$ \-\\
\<\stepcounter{ln} \theln.\> Perturb an action, $a$, in $\pi_j$ for some $\oo$
\-\\
\stepcounter{ln} \theln. {\bf Until} termination condition is met \\
\stepcounter{ln} \theln.  {\bf Return} $\pi_j$ and $Q^{\pi_j}_{\oo,a}$
\end{tabbing}}
\end{minipage}}
\caption{{\small Algorithm for  learning agent $j$'s policies  when modeled at
  level 0.}}
\label{fig:RLALG}
\end{small}
\vspace{-0.05in}
\end{figure}

Level 0 agent $j$ learns its policy while agent $i$'s actions are a part of the environment. 
As we mentioned previously, agent $j$'s level 0 model space is inclusive of the $i$'s policy space, $\Pi_i$. As the space of $i$'s policy becomes large particularly for a large planning horizon, it is intractable for $j$ to learn a policy for all $i$'s policies. In addition, considering that few of $i$'s policies are actually collaborative, we formulate a principled way to reduce the full space to those policies of $i$, denoted as $\hat{\Pi}_i$,  that could be collaborative.

\begin{figure*}
  \centering
	\includegraphics[width=0.9\textwidth]{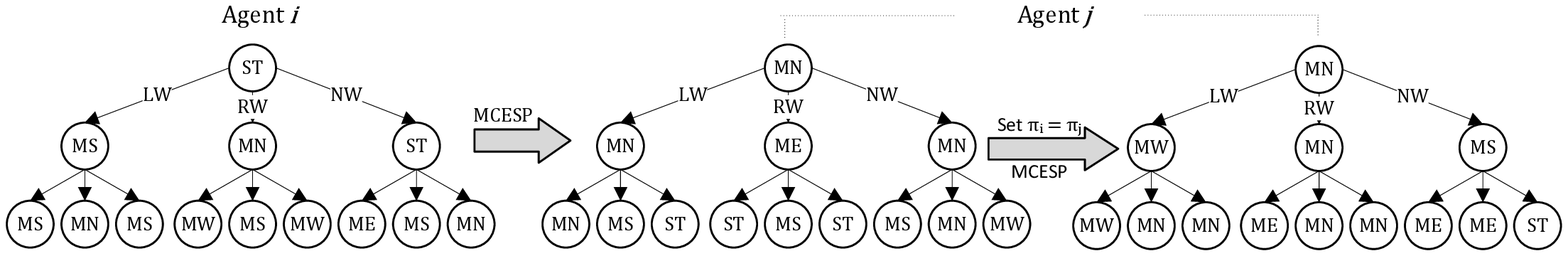}
	 \centerline{{\small $(a)$ $\pi_i$= Random initial policy}~~~~~~~~~~~~~~~~~~~{\small $(b)$ $\pi_j$=RL for Level 0 Model ($m'_{j,0}$, $\pi_i$)}~~~{\small $(c)$ $\pi'_j$=RL for Level 0 Model ($m'_{j,0}$, $\pi_j)$}}
   \caption{{\small We illustrate the principled way to generate collaborative policies for lower level agent $j$ in the 3x3 Grid domain. We start with $(a)$ a random initial policy of agent $i$ in the frame of agent $j$'s model; $(b)$ execute MCESP to generate a collaborative policy for agent $j$; and $(c)$ check whether the neighboring policy of $j$, $\pi_j'$, has a better joint utility.}}
	\label{fig:PrinWay}
\vspace{-0.05in}
\end{figure*}

We begin by picking a random initial policy of $i$ and using it in the frame of a new model of $j$. We apply generalized MCESP to this frame to obtain a candidate agent $j$'s policy, $\pi_j^1$. Next, both the initial policy of $j$ used by MCESP and $i$'s policy is set to $\pi_j^1$. MCESP then checks for the neighbors of $\pi_j^1$, which would improve on the joint utility of ($\pi_j^1$,$\pi_i$ $=$ $\pi_j^1$).
If successful, an improved neighboring policy, say $\pi_j^2$, is returned. This ensures that the joint utility of ($\pi_j^2$,$\pi_j^1$) is greater than ($\pi_j^1$,$\pi_j^1$). We continue these iterations setting $\pi_i$ as $\pi_j^2$ and using $\pi_j^2$ as the seed policy. MCESP may not improve on $\pi_j^1$ if $\pi_j^1$ is the (local) best response to $\pi_i = \pi_j^1$. Otherwise, both $\pi_j^1$ and $\pi_j^2$ are added to the set of candidate predictions of level 0 behavior of $j$. The process is restarted with a different random policy of agent $i$. We demonstrate this method on the 3$\times$3 Grid domain in Fig.~\ref{fig:PrinWay}.


\vspace{0.1in}

\subsection{Augmented I-DID Solutions}

\begin{figure}[ht!]
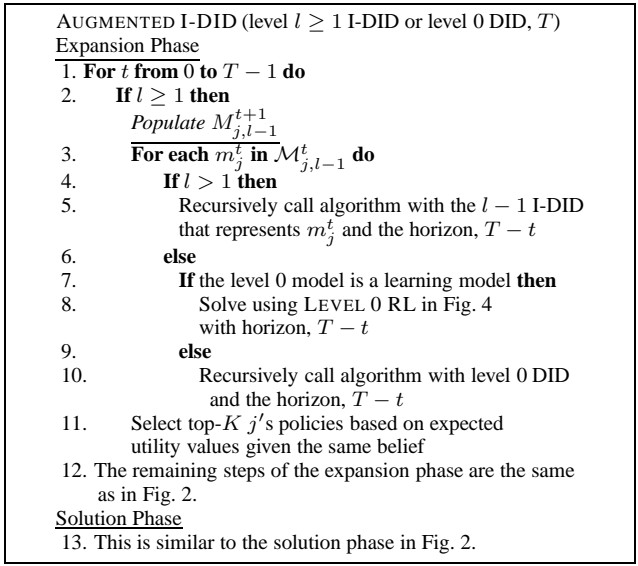

\framebox[3.3in]{
\begin{minipage}{3.2in}\small{
\begin{tabbing}
{\sc {Augmented I-DID}} (level $l \geq 1$ I-DID or level 0 DID, $T$)\\
\underline{Expansion Phase}\\
\stepcounter{ln} \theln. {\bf For}\=  $~t$ {\bf from} 0 {\bf to} $T-1$ {\bf do} \+\\
\<\stepcounter{ln} \theln.\> {\bf If}\= ~$l \geq 1$ {\bf then}\+\\
{\em \underline{Populate $M_{j,l-1}^{t+1}$}}\\
\<\<\stepcounter{ln} \theln.\>\> {\bf For}\= {\bf ~each} $m_j^t$ {\bf in} $\mathcal{M}_{j,l-1}^t$ {\bf do} \+\\
\<\<\<\stepcounter{ln} \theln.\>\>\> {\bf If}\= ~$l > 1$ {\bf
  then}\+\\
\<\<\<\<\stepcounter{ln} \theln.\>\>\>\> Recursively call algorithm with the $l-1$ I-DID\\ that represents $m_j^t$ and the horizon, $T-t$\-\\
\<\<\<\stepcounter{ln} \theln.\>\>\> {\bf else}\+\\
\<\<\<\<\stepcounter{ln} \theln.\>\>\>\> {\bf If} \= the level 0 model is a learning model {\bf then}\+\\
\<\<\<\<\<\stepcounter{ln} \theln.\>\>\>\>\> Solve using {\sc Level 0
  RL} in Fig.~\ref{fig:RLALG} \\with horizon, $T-t$\-\\
\<\<\<\<\stepcounter{ln} \theln.\>\>\>\> {\bf else}\+\\
\<\<\<\<\<\stepcounter{ln} \theln.\>\>\>\>\> Recursively call algorithm with level 0 DID\\~~and the horizon,
$T-t$ \-\-\-\\
\<\<\stepcounter{ln} \theln.\>\> Select top-$K$ $j'$s policies based on expected \\utility values given the same belief\-\-\\
\stepcounter{ln} \theln. The remaining steps of the expansion phase
are the same\\
~~~~~~~ as in Fig.~\ref{fig:ExAlgo}. \\
\underline{Solution Phase}\\
\stepcounter{ln} \theln. This is similar to the solution phase in Fig.~\ref{fig:ExAlgo}.
\end{tabbing}
}\end{minipage}
}
\caption{{\small Algorithm for solving a level $l \geq 1$ I-DID or level 0 DID
  expanded  over $T$  time steps  with $M'_{j,0}$  containing  level 0
  models that learn.}}
\label{fig:NewExAlgo}
\end{figure}

Solving augmented I-DIDs is  similar to solving the traditional I-DIDs
except that the  candidate models of the agent at level 0
may   be  learning  models.
We  show   the  revised   algorithm  in
Fig.~\ref{fig:NewExAlgo}. When the level  0 model is a learning model,
the  algorithm  invokes   the  method  {\sc  Level  0   RL}  shown  in
Fig.~\ref{fig:RLALG}.   Otherwise, it  follows the  same  procedure as
shown in  Fig.~\ref{fig:ExAlgo} to  recursively solve the  lower level
models.

While we consider a reduced space of agent $i$'s policies in a principled way, and therefore agent $j$'s learning models,  we may further reduce agent $j$'s  policy space by keeping {\em top-$K$} policies of $j$, $K$ $>$ 0, in terms of their expected utilities (line 11 in Fig.~\ref{fig:RLALG}). Observe that across models that differ in $i$'s policy and with the same initial belief, the team behavior(s) is guaranteed to generate  the largest utility in a cooperative problem. This motivates focusing on models with higher utilities. Hence, the filtering of $j$'s policy space may not compromise the quality of I-DID solutions at level 1. However, because MCESP may converge to a local optima, the resulting top-$K$ policies are not guaranteed to include $j$'s optimal collaborative policies in theory, although as our experiments demonstrate, we often obtain the optimal team behavior. As the number of optimal policies is unknown, we normally use a sufficiently large $K$ value. 

\setcounter{ln}{0}


Agent $j$'s policy space will be additionally reduced because behaviorally equivalent models -- learning and other models with identical solutions -- will be clustered~\cite{Zeng12:Exploiting}. In summary, we take several steps to limit the impact of the increase in $j$'s model space. Using a subset of $i$'s policies preempts solving all $j$'s models at level 0 while the top-$K$ technique removes possibly non-collaborative policies.

\section{Experimental Results}
Our experiments show that I-DIDs augmented with level 0 models that learn facilitate team behavior, which was previously implausible. In addition, we show the applicability of Aug. I-DIDs to ad hoc teamwork in a setting similar to the one used by Wu {\em et al.}~(\citeyear{WZCijcai11}). We empirically evaluate the performance in  {\it three}  well-known  cooperative domains involving two agents, $i$ and $j$:  $3\times3$ grid meeting ({\sf Grid})~\cite{Bernstein05:Bounded},  box-pushing ({\sf BP})~\cite{Seuken07:BoxPush}, and multi-access broadcast channel ({\sf MABC})~\cite{Hansen04:MABC}.
In the {\sf BP} domain, each agent intends to push either a small or large box into the goal region. The agents' rewards are maximum when both of them cooperatively push the large box into the goal. In the {\sf MABC} problem, nodes need to broadcast messages to each other over a channel. Only one node may broadcast at a time, otherwise a collision occurs. The nodes attempt to maximize the throughput of the channel.

\begin{table}[!ht]
  \centering
  \caption{{\upshape \small Domain Dimension and Experimental Settings}}
	\vspace{0.1in}
	{\small
    \begin{tabular}{|c|c|c|c|c|}
    \hline
		\textbf{Domain} & \textbf{T} & \textbf{$|\mathcal{M}_{j}^0|$} & \textbf{$|\hat{\Pi}_i|$} & \textbf{Dimension} \\
    \hline
		\multirow{2}[5]{*}{\textbf{Grid}} & 3     & 100 & 32 & \multirow{2}[4]{*}{$|S_j|$=9, $|S_i|$=81, $|\Omega|$=3, $|A|$=5} \\
		\cline{2-4}
          & 4     & 200 & 100 & \\	
    \hline
    \textbf{BP} & 3 & 100 & 32 & $|S|$=50, $|\Omega|$=5, $|A|$=4 \\
    \hline
    \multirow{3}[5]{*}{\textbf{MABC}} & 3     & 100 & 32 & \multirow{3}[4]{*}{$|S_j|$=2, $|S_i|$=4, $|\Omega|$=2, $|A|$=2} \\
		\cline{2-4}
          & 4     & 100 & 64 &  \\
		\cline{2-4}
          & 5     & 200 & 64 &  \\			
    \hline
    \end{tabular}%
  \label{tab:Domain}%
	}
\end{table}%

We summarize the domain properties and parameter settings of the Aug. I-DID in Table~\ref{tab:Domain}. Note that $|\mathcal{M}_{j}^0|$ is the number of initial models of agent $j$ at level 0 and $\hat{\Pi}_i$ is the subset of $i$'s policies generated using the approach mentioned earlier, allowing us to reduce the full space of $j$'s policies to those that are possibly collaborative.

\subsection{Teamwork in Finitely-Nested I-DIDs}

\noindent {\bf Experimental Settings.} We  implemented  the  algorithm  {\sc  augmented I-DID}  as  shown  in Fig.~\ref{fig:NewExAlgo}   including    an   implementation   of   the generalized  MCESP for  performing  level 0  RL.   

We demonstrate the performance of the augmented framework toward generating team behavior. We compare the expected utility of agent $i$'s policies with the values of the optimal team policies obtained using an exact algorithm, GMAA$\text{*}$-ICE, for Dec-POMDP formulations of the same problem domains~\cite{Spaan08:Multiagent}. We also compare with the values obtained by traditional I-DIDs. All I-DIDs are solved using the exact discriminative model update (DMU) method~\cite{Zeng12:Exploiting}. For both traditional and Aug. I-DIDs, we utilized $|\mathcal{M}_{j}^0|$ models at level 0 that differ in the initial beliefs or in the frame. 
We adopt two model weighting schemes: ($a$) {\bf Uniform}: all policies are uniformly weighted; ($b$) {\bf Diverse}: let policies with larger expected utility be assigned proportionally larger weights. Note that we maintain the top $K$ by expected utility (out of $|\mathcal{M}_{j}^0|$) learning and non-learning models only  while solving Aug. I-DIDs. Though the model space is significantly enlarged by the learning policies, Aug. I-DIDs become tractable when both top-$K$ and equivalence techniques are applied. 

\begin{table}[t!]
  \centering
	\caption{\upshape \small Performance comparison between the trad. I-DID, aug. I-DID, and GMAA$\text{*}$-ICE in terms of the expected utility}
	\vspace{0.1in}
  {\small
    \begin{tabular}{|c|c|cc|c|}
    \hline
    \multicolumn{1}{|c}{} & \multicolumn{3}{|c|}{\textbf{Aug. I-DID}} & \textbf{Trad I-DID} \\
    \hline
    \multicolumn{1}{|c|}{\textbf{Domain}} & \textbf{K} & \multicolumn{1}{|c}{\textbf{Uniform}} & \textbf{Diverse} & \textbf{Uniform} \\
    \hline
    \multicolumn{1}{|c|}{\multirow{2}[0]{*}{\textbf{Grid}}} & 32    & \multicolumn{1}{|c}{41.875} & 41.93 & \\
    \multicolumn{1}{|c|}{\multirow{2}[0]{*}{\textbf{(T=3)}}} & 16    & \multicolumn{1}{|c}{40.95} & 41.93 & 25.70 \\
    \multicolumn{1}{|c|}{} & 8    & \multicolumn{1}{|c}{40.95} & 41.93 &  \\
    \cline{2-5}
    \multicolumn{1}{|c|}{} & \multicolumn{4}{|c|}{\textbf {Dec-POMDP(GMAA$\text{*}$-ICE):} 43.86} \\
		  \hline
		\multicolumn{1}{|c|}{\multirow{2}[0]{*}{\textbf{Grid}}} & 100    & \multicolumn{1}{|c}{37.15} & 53.26 &  \\
    \multicolumn{1}{|c|}{\multirow{2}[0]{*}{\textbf{(T=4)}}} & 64    & \multicolumn{1}{|c}{35.33} & 53.26 & 21.55 \\
    \multicolumn{1}{|c|}{} & 32    & \multicolumn{1}{|c}{35.33} & 53.26 &  \\
    \cline{2-5}
    \multicolumn{1}{|c|}{} & \multicolumn{4}{|c|}{\textbf {Dec-POMDP(GMAA$\text{*}$-ICE):} 58.75} \\
		 \hline
    \multicolumn{1}{|c|}{\multirow{2}[0]{*}{\textbf{BP}}} & 32    & \multicolumn{1}{|c}{73.45} & 76.51 & \\
    \multicolumn{1}{|c|}{\multirow{2}[0]{*}{\textbf{(T=3)}}} & 16    & \multicolumn{1}{|c}{73.45} & 76.51 & 4.75 \\
    \multicolumn{1}{|c|}{} & 8    & \multicolumn{1}{|c}{71.36} & 76.51 &  \\
    \cline{2-5}
    \multicolumn{1}{|c|}{} & \multicolumn{4}{|c|}{\textbf {Dec-POMDP(GMAA$\text{*}$-ICE):} 85.18} \\
     \hline
    \multirow{2}[0]{*}{\textbf{MABC}} & 32   & 2.12  & 2.30  &  \\
     \multirow{2}[0]{*}{\textbf{(T=3)}}     & 16    & 2.12  & 2.30  & 1.79 \\
          & 8    & 2.12  & 2.30  &  \\
          \cline{2-5}
          & \multicolumn{4}{|c|}{\textbf {Dec-POMDP(GMAA$\text{*}$-ICE):} 2.99} \\
					  \hline
    \multirow{2}[0]{*}{\textbf{MABC}} & 64    & 3.13  & 3.17 &  \\
     \multirow{2}[0]{*}{\textbf{(T=4)}}          & 32    & 3.13  & 3.17   & 2.80 \\
          & 16    & 3.13  & 3.17   &  \\
          \cline{2-5}
          & \multicolumn{4}{|c|}{\textbf {Dec-POMDP(GMAA$\text{*}$-ICE):} 3.89} \\
					\hline
		\multirow{2}[0]{*}{\textbf{MABC}} & 64    & 4.08  & 4.16 & \\
     \multirow{2}[0]{*}{\textbf{(T=5)}}          & 32    & 3.99  & 4.16   & 3.29 \\
          & 16    & 3.99  & 4.16   &  \\
          \cline{2-5}
          & \multicolumn{4}{|c|}{\textbf {Dec-POMDP(GMAA$\text{*}$-ICE):} 4.79} \\
					 \hline
    \end{tabular}%
    }
  \label{tab:ppEU}%
\end{table}%

\noindent {\bf Performance Evaluation.} In Table~\ref{tab:ppEU}, we observe that the Aug. I-DID significantly outperforms the traditional I-DID where level 0 agent $j$ does not learn. Aug. I-DID's solutions approach the globally optimal team behavior as generated by  GMAA$\text{*}$-ICE. In cooperative games, the globally optimal solution is the pareto optimal Nash equilibrium. We observe that the larger weights on the learned policies lead to better quality $i$'s policies. This restates the importance of the augmented level 0 $j$'s models that learn. The small gap from the optimal DEC-POMDP value is due to the uncertainty over different models of $j$. 
Note that DEC-POMDPs are informed about the initial belief setting (and do not face the issue of bounded rationality) whereas, I-DIDs are not and they consider the entire candidate model space of $j$. 
{\em Furthermore, the Aug. I-DID generates the optimal team behavior identical to that provided by GMAA$\text{*}$-ICE when $i$'s belief places probability 1 on the true model of $j$, as is the case for Dec-POMDPs.} Increasing $K$ does not have a significant impact on the performance as $K$ is large enough to cover a large fraction of collaborative policies of agent $j$ including the optimal teammate.

\begin{figure}[!ht]
\begin{minipage}{3.3in}
\begin{center}
\begin{minipage}{3.3in}
\includegraphics[width=1.60in]{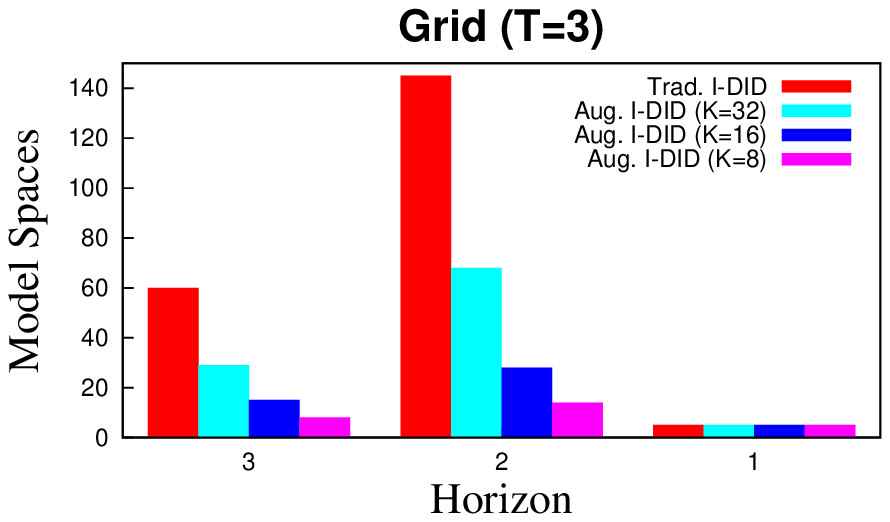}
\includegraphics[width=1.60in]{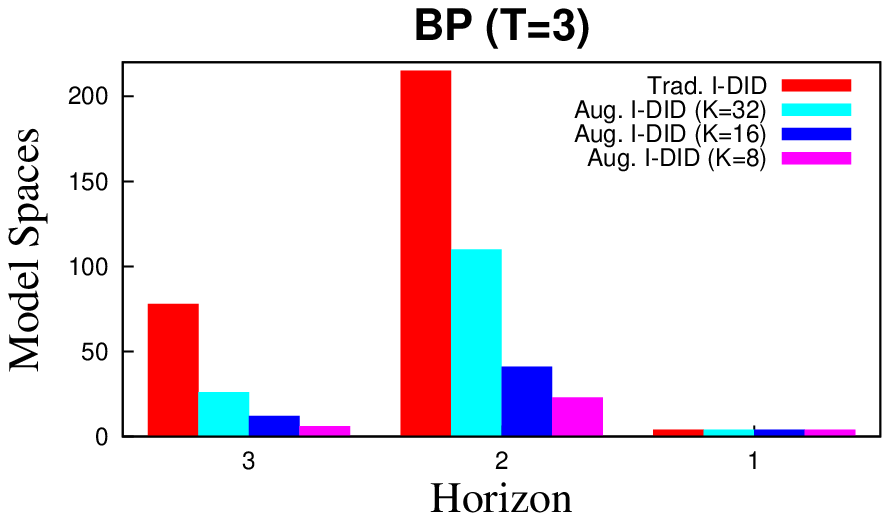}
\end{minipage}
\begin{minipage}{3.3in}
\includegraphics[width=1.60in]{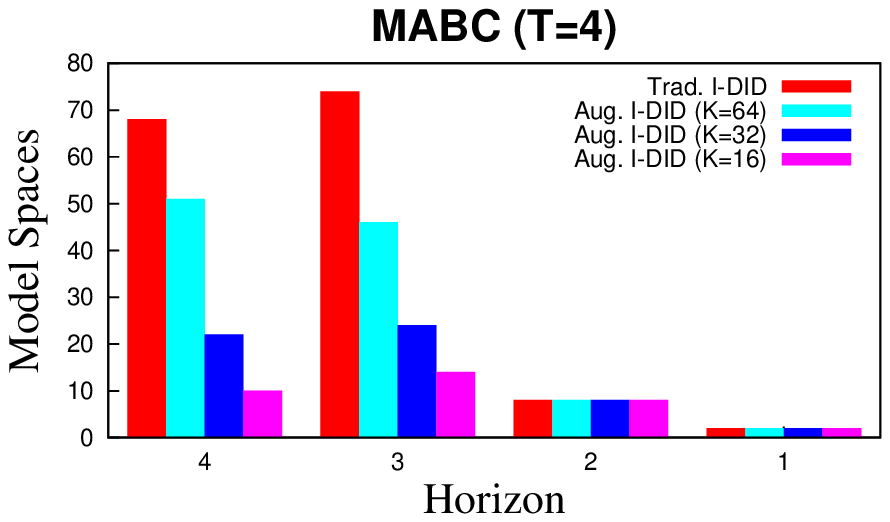}
\includegraphics[width=1.60in]{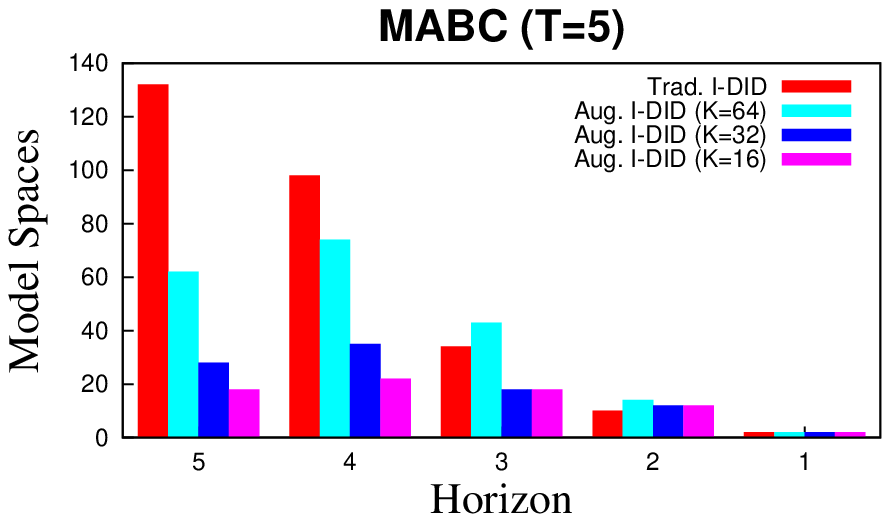}
\end{minipage}
\end{center}
\end{minipage}
\caption{{\small Top-$K$ method reduces the added solution complexity of the augmented I-DID. The complexity is dominated by the model space (number of models) in each time step.}}
\label{fig:ppMS}
\end{figure}

In Fig.~\ref{fig:ppMS}, we illustrate the reduction of model space that occurs due to smaller values of $K$, which facilitates efficiency in the solution of the Aug. I-DID. Though the augmented level 0 model space is much larger than that of its traditional counterparts, the growth in the number of models is limited due to the top-$K$ heuristic. 


\subsection{Applications to Ad Hoc Teams}

\noindent {\bf Experimental Settings.} We test the performance of the augmented I-DIDs in ad hoc applications involving different teammate types particularly when the teammates' policies are not effective in advancing the joint goal (i.e. ad hoc) and compare it with a well-known ad hoc planner, OPAT. Teammate types include: ($a$) \textbf{Random} - when the teammate plays according to a randomly generated action sequence for the entire length of the run. Some predefined random seeds are used to guarantee that each test has the same action sequences. ($b$) \textbf{Predefined} - when the teammate plays according to some predefined patterns which are sequences of random actions with some fixed repeated lengths that are randomly chosen at the beginning of each run. For example, if the action pattern is``1324'' and the repetition value is 2, the resulting action sequence will be ``11332244''. ($c$) \textbf{Optimal} - when the teammate plays rationally and adaptively. OPAT uses an optimal teammate policy for simulations, which is computed offline with the help of a generative model by value iteration. Note that OPAT in its original form assumes perfect observability of the state and joint actions. For comparison, we generalized OPAT to partially observable settings by considering observation sequences. 

Additionally, in order to speed up the generation of RL models at level 0, we implemented an approximate version of our generalized MCESP called the Sample Average Approximation (MCESP-SAA) that estimates action values by taking a fixed number of sample trajectories and then comparing the sample averages~\cite{Perk02:RLPOMDP}. We used a sample size of $n$=25 trajectories to compute the approximate value of the policy that generated them, for MCESP-SAA. We set $\alpha$=0.9, and terminate the RL (line 15 in Fig.~\ref{fig:RLALG}) if no policy changes are recommended after taking $n$ samples of the value of each observation sequence-action pair~\cite{Perk02:RLPOMDP}. We also tested with some domain-specific seed policies to investigate speedup in the convergence of MCESP. 

Simulations were run for 20 steps and the average of the cumulative rewards over 10 trials are reported for similar teammate settings for the 3 problems. We show that the augmented I-DID solution \textit{significantly} outperforms OPAT solutions in all problem domains for {\sf random} and {\sf predefined} teammates while performing comparably for {\sf optimal} ones. 

\begin{table}[t!]
  \centering
  \caption{\upshape \small Baseline Comparison with OPAT with different types of teammates. Each datapoint is the average of 10 runs.}
	\vspace{0.1in}
    \begin{tabular}{|c|c|c|}
    \hline
    \textbf{Ad Hoc Teammate } & \textbf{OPAT } & \textbf{Aug. IDID} \\
    \hline
    \multicolumn{3}{|c|}{\textbf{Grid} \textit{T=20, look-ahead=3}} \\
		\hline
		Random & 12.25 $\pm$ 1.26  & 14.2 $\pm$ 0.84 \\
		Predefined & 11.7 $\pm$ 1.63  & 16.85 $\pm$ 1.35 \\
		Optimal & 28.35 $\pm$ 2.4  & 27.96 $\pm$ 1.92 \\
		\hline
    \multicolumn{3}{|c|}{\textbf{BP} \textit{T=20, look-ahead=3}}\\
		\hline
		Random & 29.26 $\pm$ 2.17  & 36.15 $\pm$ 1.95 \\
		Predefined & 41.1 $\pm$ 1.55  & 54.43 $\pm$ 3.38 \\
		Optimal & 52.11 $\pm$ 0.48  & 59.2 $\pm$ 1.55 \\
		\hline
    \multicolumn{3}{|c|}{\textbf{MABC} \textit{T=20, look-ahead=3}} \\
		\hline
		Random & 9.68 $\pm$ 1.37 & 12.13 $\pm$ 1.08 \\
		Predefined & 12.8 $\pm$ 0.65 & 13.22 $\pm$ 0.21 \\
    Optimal & 16.64 $\pm$ 0.28 & 15.97 $\pm$ 1.31 \\
		\hline
    \end{tabular}%
  \label{tab:ppBC}%
\end{table}%

\begin{figure}[!ht]
\vspace{-0.65in}
  \centering
    \includegraphics[width=3.1in]{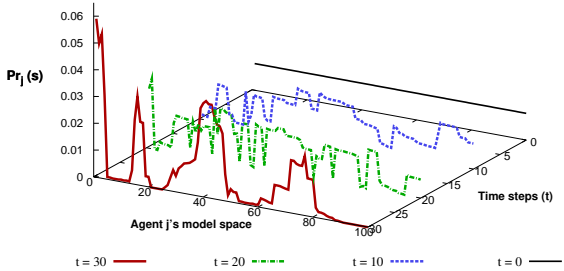}
		\vspace{-0.2in}
  \caption{{\small MABC belief updates over 30 steps showing the beliefs converging to fewer models (largest belief is that of agent $j$'s true model).}}
\label{fig:mabcBeliefs}
\end{figure}


\noindent {\bf Performance Evaluation.} Table~\ref{tab:ppBC} shows that I-DIDs \textit{significantly} outperform OPAT especially when the other agents are {\it random} or {\it predefined} types in all three problem domains (Student's t-test, $p$-value $\leq$ 0.001 for both) except when the teammate is of type {\it predefined} in the {\sf MABC} problem where the improvement over OPAT was not significant at the 0.05 level ($p$-value = 0.0676). Aug. I-DID's better performance is in part due to the sophisticated belief update that gradually increases the probability on the true model if it is present in agent $j$'s model space as shown in Fig.~\ref{fig:mabcBeliefs} for {\sf MABC}. 
As expected, both OPAT and Aug. I-DID based ad hoc agent perform better when the other agent in the team is optimal in comparison to {\it random} or {\it predefined} type. Aug. I-DIDs perform significantly better than OPAT when faced with optimal teammates for the BP domain, while the results for the other domains are similar. 

In summary, the Aug. I-DID maintains a probability distribution over different types of teammates and updates both the distribution and types over time, which differs from OPAT's focus on a single optimal behavior of teammates during planning.  Consequently, Aug. I-DIDs allow better adaptivity as examined above. Further experiments on the robustness of Aug. I-DIDs in dynamic ad hoc settings showed that agent $i$ obtained significantly better average rewards compared to OPAT ($p$-value = 0.042)  
for the setting where the other agent is of type {\it predefined} and after 15 steps is substituted by an {\it optimal} type for the remaining 15 steps in the {\sf MABC} domain.

\begin{figure}[!ht]
  \centering
    \includegraphics[width=2.5in]{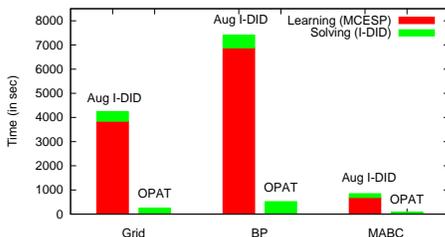}
		\vspace{-0.2in}
  \caption{{\small Timing results for augmented I-DID simulations and OPAT with an \textsc{optimal} teammate.}}
\label{fig:timing}
\end{figure}


In Fig.~\ref{fig:timing}, we show the online run times for the Aug. I-DID and generalized OPAT approaches on the three problem domains. Expectedly, OPAT takes significantly less time because it approximates the problem by solving a series of stage games modeling the other agent using a single type. In the case of Aug. I-DIDs, we observe that generating and solving the added learning models consume the major portion  of the total time. We show the learning overhead for {\sf Grid}, {\sf BP}, and the {\sf MABC} domains in red in the figure. To reduce this overhead and speed up Aug. I-DIDs, an avenue for future work is to try other RL methods in place of MCESP.

\subsection{Scalability Results} 
Although we recognize that the learning component (MCESP) is the bottleneck to scaling augmented I-DIDs for larger problems, we were successful in obtaining the optimal teammate policies using augmented I-DIDs (same as those computed by {\bf GMAA$\text{*}$-ICE}) in the $4\times4$ {\sf grid}, {\sf BP} for T=4, and {\sf MABC} for T=5. For these larger problems, we also noticed a significant improvement in the values obtained by augmented I-DIDs over their traditional counterparts as shown in Table~\ref{tab:ppEU}. In the larger $4\times4$ {\sf grid} domain for $T$=3, the optimal team value of 29.6 is achieved by the augmented I-DID compared to 19.82 obtained by solving the traditional I-DID. A better substitute for MCESP and other approximation techniques, will allow us to further scale-up augmented I-DIDs.

\section{Discussion and Conclusion}
Self-interested  individual  decision  makers  face  hierarchical  (or nested) belief systems in their multiagent planning problems. In this paper, we explicate one negative consequence of bounded rationality:  the  agent  may not  behave  as an  optimal teammate.   In  the  I-DID  framework  that  models individual  decision makers  who recursively model other  agents, we  show  that reinforcement learning integrated with the planning allows the models to produce sophisticated  policies. For the first time,  we see the principled and comprehensive  emergence  of team behavior in I-DID solutions facilitating I-DIDs' application to ad hoc team settings for which they are just naturally well-suited for. We show that integrating learning in the context of I-DIDs helps us provide a solution to a few fundamental challenges in ad hoc teamwork -- building a single autonomous agent that can plan individually in partially observable environments by adapting to different kinds of teammates while making no assumptions about its teammates' behavior or beliefs and seeking to converge to their true types. Augmented I-DIDs compare well with a standard baseline algorithm, OPAT.

While individual decision-making frameworks such as I-POMDPs and I-DIDs are thought to be well suited for non-cooperative domains, we show that they may be applied to cooperative domains as well. Integrating learning while  planning provides a  middle ground (or a bridge) between multiagent planning frameworks such as Dec-POMDPs and joint learning  for  cooperative domains~\cite{Panait05:CM}. Additionally, augmented  I-DIDs  differentiate  themselves from  other centralized cooperative frameworks by  focusing on  the behavior of  an individual agent  in a multiagent  setting. While we recognize that the introduction of learning-based models adds a significant challenge to scaling I-DIDs for larger problems,  we successfully obtained optimal teammate policies using Aug. I-DIDs in the $4\times4$ {\sf Grid} and {\sf BP} using a combination of intuitive pruning techniques. By allowing models formalized as I-DIDs or DIDs to vary in the beliefs and frames, we considered an exhaustive and general space of models during planning. The convergence of RL is not predicated on any prior assumptions about other's models. Immediate lines  of future  work involve improving the scalability of the framework, particularly  its learning component, by utilizing larger problems.


%




\bibliographystyle{abbrv}
\bibliography{AdHoc-RL-IDID-MSDM}

\end{document}